**Title:**

# Econometric applications of high-breakdown robust regression techniques

**Authors:**

Asad Zaman, *IIIE, International Islamic University of Islamabad, Pakistan*

Peter J. Rousseeuw, *Department of Math and Comp Sci, Univ Instelling Antwerpen (UIA), Antwerp,*

Mehmet Orhan, *Department of Econ and Comp Sci, Fatih University, Istanbul, Turkey*

**Abstract**

A literature search shows that robust regression techniques are rarely used in applied econometrics. We present a technique based on Rousseeuw and Van Zomeren [Journal of the American Statistical Association, 85 (1990) 633–639] that removes many of the difficulties in applying such techniques to economic data. We demonstrate the value of these techniques by re-analyzing three OLS-based regressions from the literature.

## 1. Introduction

Since it is well-established that even high quality data tend to contain outliers, and it is a rare economic data set that qualifies as 'high-quality', one would expect far greater reliance on robust regression techniques than is actually observed. Searching through ECONLIT for the term 'robust regression' led to a remarkably low total of 14 papers published in different journals most of which apply robust regression techniques to real economic data sets. We believe that this reluctance to use robust regression techniques may be due to the following factors.

- The belief that large sample sizes make robust techniques unnecessary — with enough data, one can find the truth.
- The belief that outliers can be detected simply by eye, or by looking for unusual OLS residuals, or by sensitivity analysis, obviating the need for a robust analysis.
- Existence of several 'robust regression' techniques with little guidance available as to which is appropriate.
- Unfamiliarity with interpretation of results from a robust analysis.
- Unawareness of gains available from robust analysis in real data sets.

In this paper we hope to remove some of these barriers and facilitate more routine use of a particular set of robust regression procedures, based on Rousseeuw and Van Zomeren (1990) and modern algorithms. We comment on the factors listed above, starting with the large sample size illusion.

### 1.1. *OLS and large samples*

Since OLS has a 0% breakdown value (meaning that an arbitrarily small percentage of bad observations can change the OLS coefficients to any value at all from $-\infty$ to $+\infty$), even a small proportion of deviant observations in a very large sample can cause systematic distortions in OLS estimates. For a real example, see Rousseeuw and Wagner (1994), who study distortions in earnings functions introduced by a few outliers in the form of uneducated millionaires and unemployed Ph.Ds. As another example, Knez and Ready (1997) show that the unexpected findings of size and book-to-market premia on stocks disappear if 1% of the most extreme observations are trimmed. Both these examples show the sensitivity of OLS estimates to outliers, even in large samples.

### 1.2. *Removing outliers*

In small sample sizes (as well as in large ones) OLS residuals are of little help in identifying outliers. Rousseeuw and Leroy (1987) present many real data sets in which OLS residuals fail to find any unusual data although big outliers exist. Sensitivity analysis techniques delete one observation and assess the effects of such deletions on the regression output. These were recommended by Belsley et al. (1980), and are implemented in MINITAB and many other packages. Unfortunately, when outliers are clustered, they 'mask' each other, and sensitivity analysis fails to detect such outliers. In low dimensional data sets we can resort to plotting as an effective method for locating outliers, but there do not seem to be any simple methods for visually detecting outliers in high dimensional data sets — see Rousseeuw and Van Zomeren (1990) for a discussion. OLS output, even suitably supplemented with sensitivity analysis, will not produce any signals about the presence of distortionary outliers.

### 1.3. *What is robust*?

Another factor which causes difficulty for practitioners is the presence of a large number of techniques bearing the name 'robust'. A review and comparison of strengths and weaknesses of alternative approaches is given in Chapter 5 of Zaman (1996). In this paper we propose to use a simple and intuitive criterion for robustness. We will call a procedure (for estimation and/or testing) robust if it is not altered by removing or modifying a small percentage of the data set. It is a little-known fact that most techniques labelled 'robust' in the literature do not have this property.

A popular approach is based on Huber's M-estimators, which achieve robustness in the case of location parameters. Unfortunately generalizations to regression models fail to achieve robustness. As Rousseeuw (1984) shows, regression M-estimators also have 0% breakdown value. Generalizations of M-estimators by Mallows, Schweppe, and others also fail to achieve high breakdown values.

Some studies use the Krasker–Welsch estimator to achieve robustness, as recommended by Krasker, Kuh and Welsch (1983) in their influential review article 'Estimation with dirty data and flawed models'. Although this estimator has many nice asymptotic properties, it is not robust in the sense that a small percentage of deviant observations can cause large changes in this estimator. See Zaman (1996, p. 114) and Yohai (1987) for an illustration of this failure and some discussion.

## 2. Robust regression via LTS

Methods which achieve the goal of being insensitive to changes in a small percentage of the observations have only recently been developed. Rousseeuw (1984) developed the first practical robust regression estimators (least median squares (LMS), least trimmed squares (LTS), and variants) which behave reasonably even in the presence of a large number of outliers. The gains available from these high breakdown techniques are discussed in the context of three examples of regressions, taken from economics literature. We illustrate how to identify a small subset of points, the removal of which makes a big impact on the original regression findings.

We will focus on the LTS (least trimmed squares) estimator, which has high breakdown value, in this exposition. Consider the regression model $y_t = x_t \beta + \varepsilon_t$ for $t = 1, 2, ..., T$, where $y_t$ is scalar, $x_t$ is a $1 \times K$ vector of regressors, and $\beta$ is a $K \times 1$ vector of unknown coefficients. Given any $\beta$, define the squared residual $r_t^2(\beta) = (y_t - x_t \beta)^2$, and order them in an increasing sequence: $0 \le r_{(1)}^2(\beta) \le r_{(2)}^2(\beta) \le ... \le r_{(T)}^2(\beta)$. Let α be a trimming percentage, such as 10%, for example. Then the α-trimmed LTS estimator is defined as the value of β which minimizes the following sum of squares:

$$SS(\beta) = \sum_{t=1}^{t=(1-\alpha)T} r_{(t)}^2$$

Rousseeuw (1984) showed that such estimators achieve a high breakdown value — that is, they continue to give reasonable results even in the presence of many bad observations. It is worth noting that SS(β) is a highly nonlinear, non-differentiable function with multiple local maxima. Standard algorithms for numerical maximization can easily fail to converge to global maximum, and efficient numerical optimization techniques are described in Rousseeuw and Van Driessen (1998)

A straightforward approach to robust regression is to run an LTS analysis, and flag observations having unusually large residuals. The LTS method provides an efficient way of detecting outliers, which is hard to do otherwise in a high dimensional problem. Then we can run a standard OLS regression without the outlying observations, as was proposed in Rousseeuw (1984). However, it may happen that this approach flags too many points as outliers. Removing too many points runs the risk that the resulting regression may not reflect the relationship which the econometrician desires to estimate. One way of avoiding this problem is to study the *leverage* of observations. Observations whose $x_t$ lie far away from the bulk of the $x_t$ have high leverage, in the sense that they exercise a big influence on the OLS regression coefficients. Observations whose $x_t$ lie close to the center have little leverage; even if they have high residuals they do not affect the shape of the regression relationship very much. As Rousseeuw and Van Zomeren (1990) suggest, the leverage of the *t*-th observation is calculated as the distance between $x_t$ and the center of the $x_t$ of all the observations. For this, it is very important to use a *robust* estimate for the centerpoint of the $x_t$ and a *robust* estimate for the covariance $\Sigma$ of the $x_t$. For data analysis we should at least eliminate the *bad leverage points* — these are points with high leverage which also have large LTS residuals. Such points are especially damaging to an OLS analysis. We present practical illustrations below of how to identify and eliminate outliers, and the effects of such elimination on the OLS analysis.

2.1. *Interpretation and value of robust analysis*

The next section will present a way to effectively utilize these high-breakdown techniques for data analysis in the context of a regression model. Software supporting this analysis is readily available. The present paper uses the least trimmed squares (LTS) technique, which has better statistical efficiency. Recently a fast algorithm for LTS regression was constructed (Rousseeuw and Van Driessen, 1998), which moreover can deal with very large data sets. The source code of the new program FAST-LTS is available from the website http://win-www.uia.ac.be/u/statis/index.html. The LTS is also available in S-Plus (as *ltsreg*) and in SAS/IML version 7 (as *LTS*).

The current paper employs the minimum covariance determinant (MCD) method for computing robust distances. Rousseeuw and Van Zomeren (1990) calculated robust distances with the minimum volume ellipsoid method, but the MCD is more advantageous in terms of efficiency, and it recently became available for routine use because of a new and fast algorithm (Rousseeuw and Van Driessen, 1999). The source code of the FAST-MCD program is available from the website mentioned above. This new algorithm was recently incorporated into S-Plus 4.5 (as the function *cov.mcd*) and in SAS/IML 7 (as the function *MCD*). The outputs of these programs are easy to read and evaluate.

Before proceeding it is worth noting that analysis of outliers is not a purely mechanical task. An outlier may arise for many different reasons, and different reasons may require different treatments. One has to be careful while applying the robust regression techniques.

If an outlier arises from a recording/measurement error, then eliminating it is a good solution. However, if the outlier represents a valid observation, it may point to some significant behavior falling out of the range of our model. If all outliers from the Ptolemaic circular orbits had been smoothed over or eliminated, it is possible that Copernicus would have never discovered the elliptical orbits! In any case, proper analysis of outliers requires understanding why the outlier arose, and this requires specific knowledge about the process by which the data were gathered as well as the process which is generating the data. This is also the view point of Knez and Ready (1997), who consider that the analysis of the small percentage of points which lead to aberrant results will lead to a deeper understanding of the market. Relatively mechanical analysis of the kind we will illustrate below provides input to deeper analysis, which is essential. This deeper analysis inevitably requires knowledge of the subject matter itself, and cannot be treated here.

## 3. Growth study of De Long and Summers

Our first example is a regression analysis carried out by De Long and Summers (1991) about national growth of 61 countries from all over the world from 1960 to 1985. Their main claim is that there is a strong and clear relationship between equipment investment and productivity growth. The regression equation they use and on which we will concentrate is:

$$GDP = \beta_0 + \beta_1 LFG + \beta_2 GAP + \beta_3 EQP + \beta_4 NEQ + \varepsilon \tag{1}$$

where the response variable is the GDP growth per worker (GDP) and the regressors are the constant term, labor force growth (LFG), relative GDP gap (GAP), equipment investment (EQP), and non-equipment investment (NEQ).

Table 1
Classical and robust regression results for De Long and Summers data on national growth

| Var. | OLS | | | | ROBUST | | | |
|---|---|---|---|---|---|---|---|---|
| | Coeff. | S.E. | $t$-value | $P$-value | Coeff. | S.E. | $t$-value | $P$-value |
| Con | 20.014 | 0.0103 | 21.391 | 0.170 | 20.022 | 0.0934 | 22.379 | 0.021 |
| LFG | 20.030 | 0.1984 | 20.150 | 0.881 | 0.045 | 0.1771 | 0.253 | 0.802 |
| GAP | 0.020 | 0.0092 | 2.208 | 0.031 | 0.025 | 0.0082 | 2.982 | 0.004 |
| EQP | 0.265 | 0.0653 | 4.064 | 0.000 | 0.283 | 0.0581 | 4.859 | 0.000 |
| NEQ | 0.062 | 0.0348 | 1.791 | 0.079 | 0.085 | 0.0314 | 2.703 | 0.009 |
| | $R^2$ | 0.34 | $F$-value | 7.175 | $R^2$ | 0.44 | $F$-value | 11.000 |

Although the authors selected different time periods and sets of countries — which led them to carry out many different regressions — we focused on the main regression, stated above, where the time period is the longest, from 1960 to 1985, and all 61 countries are included.

Table 1 displays regression statistics for both the non-robust technique applied in the original paper, OLS, and the robust regression technique (ROBUST) that detects and eliminates the vertical outlier in the way explained in the previous section. Robust distances were obtained, as well as the standardized LTS residuals. Zambia appeared as the country with a very high standardized LTS residual, and is excluded from the group of observations — the LTS residual of Zambia was — 5.41, which indicates that Zambia is a very significant outlier. OLS was run with and without this point, and the results are reported in Table 1

Our robust regression technique shows that Zambia is an outlier. Eliminating just this one point has a substantial effect on the OLS analysis of De Long and Summers. The non-equipment investment (and also the constant term) appear insignificant in the original analysis, but our robust analysis reveal them to be significant. There is also a substantial improvement in the fit, indicated by the $F$-statistic, the $R^2$, and the standard error of the regression.

## 4. Solow model applied to the OECD countries

Nonneman and Vanhoudt (1996) introduce human capital to the Mankiw et al. (1992) study on the augmented Solow model. The augmented Solow model suggests

$$\ln(Y_t / Y_0) = \beta_0 + \beta_1 \ln(Y_0) + \beta_2 \ln(S_k) + \beta_3 \ln(N) + \varepsilon \qquad (2)$$

where $Y$ is the real GDP per capita of working age, $S_k$ is the average annual ratio of domestic investment to real GDP, and $N$ is annual population growth plus 5%. The data set contains 22 countries. The original paper tries different versions of the more general equation, and this is an arbitrary one selected among them.

It is important to observe that it is not at all easy to detect outliers in multidimensional data sets. For example, a study of the OLS residuals reveals no outliers in the present data. But there are five points with standardized LTS residuals bigger than 2.5 standard deviations in absolute value. These are Canada, Japan, Norway, Turkey, and New Zealand. Since Japan and Norway are not leverage points, we retain them in the data set. Canada, Turkey, and New Zealand have standardized LTS residuals of 4.21, —6.14 and —3.16 and robust distances 7.25, 9.36, and 5.98 which together say that these are bad leverage points. We obtain different results reported in Table 2 when these countries are removed from the data set.

Table 2
Regression results for Solow's growth model

| Var. | OLS | | | | ROBUST | | | |
|---|---|---|---|---|---|---|---|---|
| | Coeff. | S.E. | $t$-value | $P$-value | Coeff. | S.E. | $t$-value | $P$-value |
| Const. | 2.98 | 1.02 | 2.91 | 0.009 | 4.71 | 1.17 | 4.04 | 0.001 |
| $\ln(Y_0)$ | 20.34 | 0.06 | 26.07 | 0.000 | 20.41 | 0.05 | 27.60 | 0.000 |
| $\ln(S_k)$ | 0.65 | 0.20 | 3.22 | 0.005 | 0.52 | 0.18 | 2.90 | 0.011 |
| $\ln(N)$ | 20.57 | 0.29 | 21.97 | 0.064 | 20.12 | 0.35 | 20.35 | 0.729 |
| | $R^2$ | 0.75 | $F$-value | 17.7 | $R^2$ | 0.83 | $F$-value | 25.2 |

Elimination of these three countries has a large effect on the regression. As before, the fit is improved. Equally important, the population growth variable ln (*N*), which is borderline significant in the original equation, appears insignificant in the robust regression.

## 5. Benderly and Zwick's return data

Benderly and Zwick (1985) aim to explain the yearly return on USA common stocks by output growth and inflation, over the 1954–1981 period. Return on stocks is measured using the Ibbotson–Sinquefeld data base, output growth is proxied by the percentage change in real GNP, and inflation is measured by the deflator for personal consumption expenditures.

The regression equation concentrated on is:

$$R_t = \beta_0 + \beta_1 G_t + \beta_2 I_t + \varepsilon_t \quad (3)$$

where *R*, *G*, and *I* stand for return on common stocks, output growth, and inflation, respectively. A standard OLS analysis reveals no outliers among these 28 years. However, a robust LTS analysis reveals that two of the observations have standardized residuals just outside of the 2.5 standard deviations tolerance band. In addition, these years have slightly high robust distances; hence they are mild leverage points. Indeed, the robust distances and the standardized LTS residuals are 3.65 and 2.60 for 1979, whereas they are 3.55 and 2.82 for 1980. OLS was run with and without these years, and the regression results for both OLS and ROBUST are displayed in Table 3.

The OLS results suggest that inflation is not significant; but deleting just the two moderate outliers alters this result, and inflation becomes significant in the robust analysis. As usual there is substantial improvement in the fit. One would guess that the OLS analysis is misleading regarding the non-significance of inflation in the original paper.

Table 3
Benderly and Zwick's data for return on common stocks

| Var. | OLS | | | ROBUST | | |
|---|---|---|---|---|---|---|
| | Coeff. | S.E. | $t$-value | Coeff. | S.E. | $t$-value |
| Const. | −3.59 | 8.58 | −0.42 | 3.26 | 7.82 | 0.42 |
| Growth | 4.78 | 1.37 | 3.49 | 4.31 | 1.22 | 3.54 |
| Inflation | −1.05 | 1.15 | −0.91 | −2.97 | 1.16 | −2.56 |
| $R^2$ and $F$ | 0.56 | 10.5 | | 0.65 | 21.0 | |

## 6. Concluding remarks

We have applied high breakdown robust regression techniques to several linear regression models found in the recent economic literature. In each case we observed that deleting a few outlying points has a big impact on the regression results. The fit often improves and the significance of the regressors changes, with implications for theory. The application of this technique is strongly recommended, even for simple regression, to avoid the misleading effects of bad leverage points.

The classification of data into groups of regular observations, bad leverage points, good leverage points, and vertical outliers, offers the researcher important diagnostic insight in the data. The good leverage points are very valuable, because they allow to obtain the regression fit with high precision. The bad leverage points pull the OLS regression fit in the wrong direction. The vertical outliers are also capable of strongly affecting the OLS regression fit: recall that the removal of just one such significant vertical outlier — belonging to Zambia — led to a drastic change of the regression results in De Long and Summers' data set.

Although high breakdown robust regression techniques still take somewhat more computation time than plain OLS regression, their benefits more than outweigh the cost of applying them.